\documentclass[sigconf,authorversion]{acmart}
\usepackage{multirow}
\usepackage{todonotes}

\AtBeginDocument{%
  }

\copyrightyear{2023}
\acmYear{2023}
\setcopyright{acmlicensed}\acmConference[ARES 2023]{The 18th International Conference on Availability, Reliability and Security}{August 29-September 1, 2023}{Benevento, Italy}
\acmBooktitle{The 18th International Conference on Availability, Reliability and Security (ARES 2023), August 29-September 1, 2023, Benevento, Italy}
\acmPrice{15.00}
\acmDOI{10.1145/3600160.3605064}
\acmISBN{979-8-4007-0772-8/23/08}

\begin{document}

\title[GDPR Data Requests]{Needle in the Haystack: Analyzing the Right of Access According to GDPR Article 15 Five Years after the Implementation}

\author{Daniela Pöhn}
\orcid{0000-0002-6373-3637}
\author{Niklas Mörsdorf}
\author{Wolfgang Hommel}
\email{[firstname.lastname]@unibw.de}
\affiliation{%
  \institution{Universität der Bundeswehr München, RI CODE}
  \city{Neubiberg}
  \country{Germany}
}

\renewcommand{\shortauthors}{Pöhn et al.}

\begin{abstract}
The General Data Protection Regulation (GDPR) was implemented in 2018 to strengthen and harmonize the data protection of individuals within the European Union. One key aspect is Article 15, which gives individuals the right to access their personal data in an understandable format. Organizations offering services to Europeans had five years' time to optimize their processes and functions to comply with Article 15. This study aims to explore the process of submitting and receiving the responses of organizations to GDPR Article 15 requests. A quantitative analysis obtains data from various websites to understand the level of conformity, the data received, and the challenges faced by individuals who request their data. The study differentiates organizations operating worldwide and in Germany, browser website- and app-based usage, and different types of websites. Thereby, we conclude that some websites still compile the data manually, resulting in longer waiting times. A few exceptions did not respond with any data or deliver machine-readable data (GDRP Article 20). The findings of the study additionally reveal ten patterns individuals face when requesting and accessing their data.
\end{abstract}




\maketitle

\section{Introduction}

The General Data Protection Regulation (GDPR) was implemented in May 2018 with the goal of strengthening and harmonizing data protection for individuals within the European Union (EU). One of the key provisions of the GDPR is Article 15. This right allows individuals (i.\,e., data subjects) to request information about how their personal data is being processed and used, and to obtain a copy of their data from organizations. The information needs to be provided within a month, free of charge, and in an understandable way and format. If individuals find incorrect information, they can demand a change. Since its implementation, organizations have had five years to improve their processes to comply with the regulation.

This quantitative study aims to explore the process of requesingt data from organizations and receiving their responses with the differentiations of \emph{a)} smaller (Germany-based) and bigger (worldwide-operating) organizations, \emph{b)} browser website- and app-based usage, \emph{c)} organizations within and outside the EU, and \emph{d)} different types of websites (social media, e-commerce, others). Through a quantitative analysis of data obtained from websites, the study seeks to understand the level of conformity with Article 15 requests and the data received in response. In addition, this study will examine patterns in relation to GDPR Article 15.

We shed light on (RQ1) conformity with GDPR Articles 15, taking the differentiations  \emph{a)}--\emph{d)} into account, (RQ2) handling of Article 15 requests (responses, time, format, understandability vs. machine-readable data (GDPR Article 20 (1))), (RQ3) balance between obligation to provide access and identification, and (RQ4) any pattern derived from the results. The contribution of this paper is to provide an in-depth understanding of the implementation of GDPR Article 15 through a quantitative analysis, including \emph{a)}--\emph{d)}.

The remainder of the paper is as follows: First, the methodology of the study is outlined. This is followed in Section~\ref{sec:evaluation} by the results and their evaluation. Based on the results, dark patterns are identified in Section~\ref{sec:darkpattern}. The results gained are discussed in Section~\ref{sec:discussion} and compared with the literature in Section~\ref{sec:sota}, before Section~\ref{sec:conclusion} provides a conclusion and outlines future work.

\section{Methodology}
\label{sec:methodology}

In order to answer (RQ1) and (RQ2), we apply a qualitative research design. The websites are mainly selected based on their ranking on the former Alexa Top 50 list~\cite{alexa}. Websites without English versions are excluded; multiple websites from one organization are merged. Hence, we receive 21 websites. We exclude Telegram due to the non-functioning of the GDPR bot and PayPal, as they require a copy of an ID. In order to see the difference between local websites, we choose eight websites from a list of German websites~\cite{ger}, which reach a smaller audience than the other international websites. An overview is given in Table~\ref{tab:accounts}. The selected 27 websites are distributed between browser-based and app-based accounts. We assume that, other than the workflow for requesting and receiving data requests, no difference should be found. To obtain data, accounts for an actual or synthetic user are created. In order to establish a synthetic user, a user story with a corresponding SIM card and an artificial intelligence (AI)-based picture is developed. We use the actual user if proofs are likely to be requested. 

We look for GDPR Article 15, 12 (1)-(6), and, if additionally provided, 20 (1) (machine-readable data) request possibilities. If the GDPR requests are presumably automatically answered (i.\,e., if integrated functionality for requests is available), then a first request is sent before any interactions start after registration. Afterward, these websites are manually used like a normal user would (browsing, liking, commenting, and sharing data), without actually shopping for goods. The actions are documented. After a month of regular usage, another request is sent. This request is used for the evaluation. If the time span -- based on the previous response times -- allows for a third request within the study's time frame, then it is sent after another month. For email requests (presumably manually answered requests), the template provided by~\cite{datarequests} with different language versions is applied.

The collected data (data itself, format, and explanations) is analyzed using a thematic analysis approach. This involves testing for statistical significance and identifying common themes and patterns in the requests, responses, and data. The analysis is conducted by the authors and reviewed by the respective coauthors to ensure reliability and validity. We use our documentation and the data exported to answer (RQ3) and (RQ4). For (RQ3), we compare the identification required to send a request and receive the data. To answer (RQ4), we note all observed issues. If the issue appears several times, it is added as a pattern.
The data is anonymized to protect the privacy of the authors. The sample size of 27 websites may not be representative. In consequence, the findings may not be generalizable to other websites, but provide the first insights. This approach was selected due to the manual steps. The study was conducted between September 2022 and March 2023. As the implementation of the GDPR continues, the experience may change.

\begin{table}[!htpb]
\caption{Accounts used for the study}
\label{tab:accounts}
\begin{tabular}{lp{1.5cm}p{1.8cm}p{2.3cm}}
\toprule
   \textbf{Location}                        & \textbf{Type}              & \textbf{Browser}                                                              & \textbf{App}                                                                                                                         \\ \midrule
\multirow{10}{*}{Worldwide} & Social \newline Media & \begin{tabular}[c]{@{}l@{}}Twitter~\cite{twitter}\\Instagram~\cite{instagram}\end{tabular}       & \begin{tabular}[c]{@{}l@{}}Twitter~\cite{twitter}\\Instagram~\cite{instagram}\\WhatsApp~\cite{whatsapp}\\TikTok~\cite{tiktok}\\Snapchat~\cite{snapchat}\\ Threema~\cite{threema}\\ Signal~\cite{signal}\\ Skype~\cite{skype}\\Facebook~\cite{facebook}\\Reddit~\cite{reddit}\end{tabular} \\ \cmidrule{2-4}
                           & E-Commerce   & Amazon~\cite{amazon}                                                              & \begin{tabular}[c]{@{}l@{}}Amazon~\cite{amazon}\\AliExpress~\cite{ali}\\eBay~\cite{ebay}\end{tabular}                                                                    \\ \cmidrule{2-4}
                           & Others       & \begin{tabular}[c]{@{}l@{}}Google~\cite{google}\\Dropbox~\cite{dropbox}\\Microsoft~\cite{microsoft}\end{tabular} & \begin{tabular}[c]{@{}l@{}}Google~\cite{google}\\Microsoft~\cite{microsoft}\\Yahoo~\cite{yahoo}\\Wikipedia~\cite{wiki}\\Zoom~\cite{zoom}\\Twitch~\cite{twitch}\end{tabular}              \\ \midrule
\multirow{4}{*}{Germany}                    & E-Commerce   & -                                                                    & \begin{tabular}[c]{@{}l@{}}eBay Klein-\\anzeigen~\cite{klein}\\Otto~\cite{otto}\\Zalando~\cite{zalando}\end{tabular}                                              \\ \cmidrule{2-4}
                           & Others       & \begin{tabular}[c]{@{}l@{}}t-online.de~\cite{tonline}\\BILD.de~\cite{bild}\end{tabular}              & \begin{tabular}[c]{@{}l@{}}BILD.de~\cite{bild}\\Chefkoch~\cite{chefkoch}\\ARD Mediathek~\cite{ard}
\end{tabular}                                                                     \\ \bottomrule
\end{tabular}
\end{table}

\section{Evaluation}
\label{sec:evaluation}

Table~\ref{tab:requests} provides an overview of the key facts based on the sorting of Table~\ref{tab:accounts}. Here, \checkmark  shows yes,  \texttimes  no, and $o$ partly fulfillment.

\subsection{Process of Requesting Data}
\label{sec:process}

The process of requesting data consists of the steps to take for the request itself, which may include some sort of verification, the waiting time required to gather the data, the notification, and the data download, which again may require verification.

The request is either sent via a website (browser- and app-based), such as a form or a specific button, chat (for some messenger apps), or via email after 5.7 steps on average (std. 2.4). In most cases, the user must be logged in to request the data. An additional layer of security is partly added (26.3\%) with a password or a verification email with a link, which was not received in one case. If email requests are applied, then the user needs to include their user information.

\begin{description}
\item[Web page:] The number of steps varies from three up to ten (avg. 6.5, std. 2.4). Some providers show a privacy dashboard with the main information beforehand or instead.
\item[Chat:] A specific message is sent to a predefined contact. The number of steps is low, if the guideline is correct.
\item[Email:] Few steps (avg. 4.1, std. 1.5) are required to find the corresponding email address, which is typically part of the privacy statement, but might be found somewhere else as well. An email typically indicates manual processing.
\end{description}

\begin{figure}[!htpb]
\centering
\includegraphics[width=\linewidth]{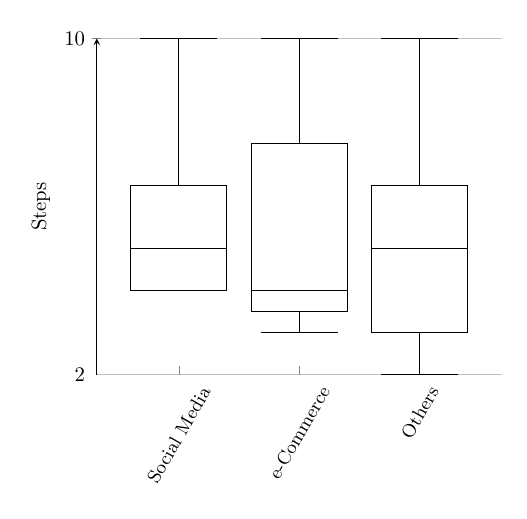}
\caption{Number of steps for the request depending on the category of service}
\Description[Number of steps for the request depending on the category of websites]{A comparison of the number of steps depending on the category of websites shows more variation for e-commerce and fewer minimum steps for others.}
\label{fig:steps}
\end{figure}

Based on the categories of the websites, we find the following, as shown in Figure~\ref{fig:steps}. Social media accounts have less variation, but require a higher minimum number of steps. Websites in the category others start with fewer steps, whereas the variation in the category e-commerce is the highest, although the average is lower.

\begin{figure}[!htpb]
\centering
\includegraphics[width=\linewidth]{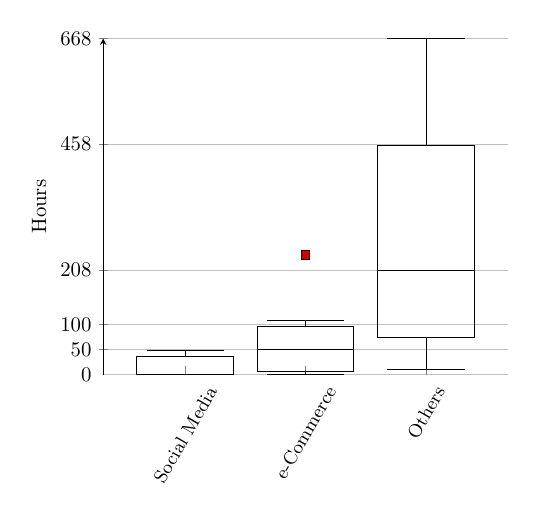}
\caption{Duration of processing depending on the category of website}
\label{fig:time}
\Description[Duration of processing depending on the category of website]{A comparison of the duration of processing depending on the category of website shows that websites in the category others typically require more time for processing.}
\end{figure}

The location may differ from browser-based website to smartphone app and between the different app versions. Hence, guidelines may not work. This was the case for Twitter. In addition, the specific app may not provide access to the request page or function, but others or the web version, see, for example, Amazon Music app. In addition, we noticed broken and almost hidden links and email addresses, for example, at Zoom and eBay Kleinanzeigen. The download was ready within one minute (automated function) to one month. On average, it took 156.7 hours. There was no difference noticed between the browser-based websites, chat, and smartphone apps. This is in contrast to the built-in function compared with email cases, see Section~\ref{sec:email}. As the email category correlates with the website category of others, we also notice a difference within the website categories, as shown in Figure~\ref{fig:time}.

If the user makes use of the website in a web browser, then a notification is sent via email to the user that their data is ready. For in-app usage, this email notification is found in 76\% of cases. It is noticeable that these 24\% only in-app notifications are not shown in the smartphone notifications of the test phone, hence, the download can be missed. All in all, 63\% of websites provided data, 59.3\% did not or not according to GDPR Article 15. The difference between data and data according to GDPR (i.\,a., understandable) is 0.04 (p-value). The numbers may change when taking other information, such as court decisions, into account. Either the data is sent via email (100\% of email cases) or a web page/chat is applied (100\% of the other cases). The data might be secured (30\%) with an extra password, token, or similar. Mostly, the user is logged in. Nevertheless, some problems occurred as the captcha was too fast (one time), the token was sent late via SMS (one time), and the download link expired faster than the set time frame (one time). If the download was provided via a browser-based website or smartphone app, then a certain time frame (48 hours to 30 days) was set to retrieve it.

\begin{table*}[!htpb]
\caption{Overview of the requests}
\label{tab:requests}
{\def\arraystretch{1.3}
\begin{tabular}{p{2cm}p{2cm}p{2cm}p{4cm}p{4.2cm}p{1.5cm}}
\toprule
\textbf{Platform}  & \textbf{Place of \newline Request} & \textbf{Place of \newline Answer} & \textbf{Format of Answer}                              & \textbf{Comment}                       & \textbf{Answered} \\  \midrule
Twitter            & Web page                  & Email, \newline web page / app      & EOT, GIF, HTML, ICO, JPG, JS, PNG, SVG, TTF, TXT, WOFF &   User-friendly      &                               \checkmark \\
Instagram          & Web page                  & App                      & JSON, TXT, JPEG  or HTML, JPEG               & Either user-friendly or machine-readable                                       &     \checkmark                           \\
WhatsApp           & Chat                      & App                      & HTML, JSON, PNG, TXT                                   &  Both                              &         \checkmark                       \\
TikTok             & Web page                  & App                      & TXT or JSON                                                   &   Machine-readable            &       \checkmark                         \\
Snapshat           & Web page                  & App                      & HTML, JSON, JPEG                                       &    Both                                    &     \checkmark                           \\
Threema            & Chat                      & App                      & JSON                                                   & Did not work in test smartphone        &    \checkmark                            \\
Signal             & None                      & -                        & -                                                      &   -                                & \texttimes
                            \\
Skype              & Web page                  & -                        & -                                                      & -                                       & \texttimes                            \\
Facebook           & Web page                  & Email, \newline web page / app      & HTML/JSON, JPEG, PNG, TXT                                   &   Either user-friendly or machine-readable                                     &     \checkmark                           \\
Amazon             & Web page                  & Email, \newline web page / app      & CSV, EML, JSON, PDF, TXT                               &  Machine-readable      &         \checkmark                       \\
AliExpress            & Web page / app                 & Web page / app                        & JPG, TXT                                                      &  Machine-readable                                    & \checkmark                            \\
Google             & Web page                  & Email                    & CSV, HTML, ICS, JSON, MP3, PDF, TXT, VCF               &  Both partly        &    \checkmark                            \\
Microsoft          & Several web pages         & -                        & -                                                      &  -                                       & $o$                           \\
Yahoo              & Web page / app                     &  Web page / app                       & CSV, JSON, PNG, TXT                                                      & Both       & \checkmark                            \\
Dropbox            & Email                     & Email                    & PDF                                                    &  User-friendly                                      &    \checkmark                            \\
Wikipedia          & Email                     & Email                    & -                                                      & Forwarded to user settings page        & \texttimes                            \\
Zoom               & Formular                  & Email                    & Email text                                             & Only general information about privacy & \texttimes                            \\
Reddit             & Email                     & App                      & -                                                      & Expired before set time frame           & \texttimes                            \\
Twitch             & Web page                  & Email                    & CSV, JSON                                              &  Machine-readable                                      &   \checkmark                             \\
eBay               & Web page                  & Email                    & HTML                                                   &     User-friendly                                   &    \checkmark                            \\
eBay Kleinanzeigen & Email                     & Email                    & HTML, XLSX                                                      & Obviously does not contain all data  & \checkmark                            \\
Otto               & Web page / Email          & -                        & -                                                      & Account got deactivated                & \texttimes                            \\
Zalando            & Web page                  & Email, \newline web page / app      & CSV, HTML                                              &   Both                                 &   \checkmark                             \\
BILD.de               & Email                     & Email                    & PDF                                                    &     User-friendly                                   & \checkmark                            \\
Chefkoch           & Email                     & Email                    & PDF                                                    &  User-friendly                                      &    \checkmark                       \\
t-online.de           & Email                     & -                        & -                                                      &  -                                      & \texttimes                            \\
ARD \newline Mediathek      & Email                     & Email                    & Email text                                             & -                                  & $o$             \\       \bottomrule
\end{tabular}
}
\end{table*}

\subsection{Data Analysis}
\label{sec:analysis}

\subsubsection{Analysis based on GDPR}
\label{sec:compliance}

\begin{figure}[!htpb]
\centering
\includegraphics[width=\linewidth]{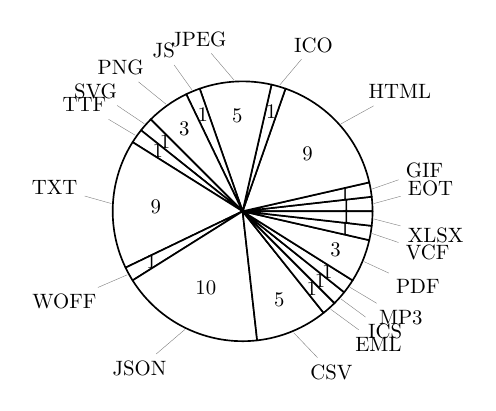}
\caption{Data formats found in the exports}
\label{fig:formats}
\end{figure}

When analyzing conformity with GDPR, we especially focus on Article 15 (request of data), but take Articles 5 (1) (i.\,a., minimization), 12 (1)-(6) (i.\,a., transparent and understandable), and 20 (1) (machine-readable) into consideration. As noted in Table~\ref{tab:requests}, not all websites answered our requests. 6/27 (22.22\%) did not answer at all, whereas three answers were not as expected (too little information or too early expired data download). In addition, we had problems downloading the second part of the data from Google (``too many requests'') and one download from Microsoft contained zero data. In most cases, it is no problem to request new data right afterward. With manual processing, it takes longer. In addition, BILD.de and Reddit reduce the number of requests to one per month. The answers included an average of 3.1 different data formats (std. 2.6; var. 6.9, see Figure~\ref{fig:formats}). The most commonly used formats are JavaScript Object Notation (JSON), HyperText Markup Language (HTML), and TXT. This is followed by Joint Photographic Experts Group (JPEG) and comma-separated values (CSV). Other formats, such as Portable Document Format (PDF), iCalendar (ICS), MP3, and vCard (VCF) are less often found.

\paragraph{Understandable and machine-readable}

\begin{figure}[!htpb]
\centering
\includegraphics[width=\linewidth]{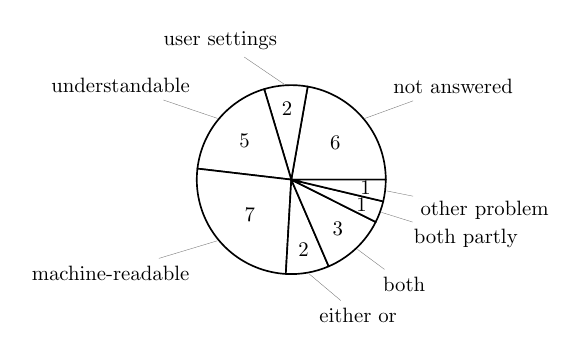}
\caption{Summary of the answered}
\label{fig:answered}
\end{figure}

Two relevant articles on GDPR are Articles 12 (1) (understandable data) and 20 (1) (machine-readable data), which do not have to be complied with at the same time. However, we found no specific functionality for Article 20 in our study. Several websites fulfill either of them (25.93\% machine-readable, 18.51\% understandable), whereas only 11.11\% provide both and one both partly. 7.4\% let the user choose between the options. Figure~\ref{fig:answered} summarizes the fulfillment of understandable and machine-readable data. Either CSV/JSON or HTML/TXT is provided, along with other file formats. The results are significantly better for those providing integrated request functions.

\paragraph{Completeness, accuracy, and data minimization}

Article 5 outlines the principles of data handling, including transparency, data minimization, and accuracy, which implies completeness. As we documented our inputs, we could verify the inclusion/exclusion and accuracy of the data exports based on them. If we notice data that might not be necessary, we write it down. In the following, we go through each website and summarize our findings.

\begin{description}
\item[AliExpress:] Neither saved products nor search history.
\item[Amazon:] Seems complete based on our documentation.
\item[BILD.de:] As the data is, according to BILD.de, the same after one month of usage, neither favorites, read and saved articles, nor search- and browser history are presumably stored. According to the privacy statement, they log the Internet Protocol (IP) address and cookies during a browser session. With the use of the app, this information is probably also stored. The privacy statement further enumerates the frequency and duration of usage, views of articles and pictures, the play of video and audio files, region of usage, and actions, such as comments and shares of links. We cannot find this information in our data export.
\item[Chefkoch:] In the data export, we cannot find stored recipes, recipe books, or the search history. Taking a look at the privacy statement, we notice that the Android app applies Google Drive as the default for auto-backup of the app with its settings, including the missing information. The synchronization with Google Drive can be deactivated within the Google Drive app, which implies that there are no alternative ways in the Chefkoch app. We, though, cannot find any hint in the data export about this. In addition, a user token is sent to Algolia for a personalized search, which we cannot recognize in the data export.
\item[Dropbox:] Does not include information about uploaded data.
\item[eBay:] Seems complete based on our documentation.
\item[eBay Kleinanzeigen:] Neither stored products nor searches.
\item[Facebook:] Seems complete based on our documentation. We find activities outside of Facebook, namely AliExpress, TikTok, Microsoft Edge, Zalando, eBay Kleinanzeigen, Twitch, and Dropbox, in the data export of the synthetic user.
\item[Google:] Although we used Google Chrome on the smartphone (synthetic user), we cannot find any corresponding data. The same problem appears for emails. Searches are included as such without the actual search term. In contrast, we also find advertisements in other apps, i.\,e., eBay Kleinanzeigen, Chefkoch, and BILD.de, in the data export of the synthetic user. For the actual user, browser and search history are included, and the emails are exported in MBOX format.
\item[Instagram:] We cannot find comments and likes. Also, we applied a filter for one picture, which is neither included in the original state nor in its metadata.
\item[Twitch:] Seems complete based on our documentation.
\item[Snapchat:] Seems complete based on our documentation.
\item[TikTok:] Seems complete based on our documentation. We find network type, carrier, device system and model, IP address, and more data.
\item[Threema:] Seems to include the most important information.
\item[Twitter:] Seems complete based on our documentation. It includes information about the IP addresses and device types.
\item[WhatsApp:] We cannot find messages and contacts.
\item[Yahoo:] The written and received emails cannot be found in the data export. The numbers 1 and 4 of four personalized ad folders are included. Although this app was almost never used, it includes lots of data about the device, including IP geolocation and individual device names.
\item[Zalando:] Seems complete based on our documentation and includes information about the rights according to GDPR.
\end{description}

As noticed with BILD.de and Chefkoch, the privacy statement might reveal more missing information. Information about the authentication method(s) is typically not included.  In addition, we found indicators for tracking or risk-based authentication. In the following, we explain the problems we had and our procedure with the following websites.

\begin{description}
\item[ARD Mediathek:] The email request sent to Südwestrundfunk, i.\,e., the coordinating data protection office, first caused confusion, which was solved by an email exchange. The answer followed in plain text with the email address, an explanation of a user ID, and information about adulthood.
\item[Microsoft:] Microsoft forwards the user to a privacy dashboard. From there, the user can choose which service they want to request their data from. The requests are handled differently, from web forms to integrated functionality and email.
While the natural user received two of the three answers (the third answer contained zero data), the synthetic user received none of the two possible answers before the submission of the paper. As the majority of the requests were negative, we count them as neutral to negative.
\item[Reddit:] In the case of Reddit, the link for the download expired before the set date. Hence, a second download was requested, which was denied as it was the second request within the month (the same duration as the download timeframe). The authors could not access the third download requested as the link expired right away.
\item[Otto:] The artificial account was deactivated after one month, before sending the GDPR request, while still obtaining advertisement emails. This might be due to the potentially received rating information.
\item[Signal:] Signal has the label of a GDPR-friendly and conforming app. Although the list of contacts on the smartphone is optionally read, it is only stored in-app. According to Signal~\cite{siggdpr}, as everything is end-to-end encrypted, the Signal service has no access to any of this (message, pictures, files, or other content stored locally) data.  While it is possible to export data from Signal, there is no foreseen workflow to request data according to GDPR Article 15. To decrypt the data from the backup/export, further tools such as signal-backup-decode for Linux~\cite{backup} are required. The data then includes attachments, avatars, preferences, stickers, and the \texttt{signal\_backup.db} with all the data from the app. This might be counted as a GDPR export of data, but nonetheless, the data is not presented in an understandable way.
\item[Skype:] The request was not answered during the study.
\item[t-online.de:] As the responsible person was unsure if we had any data at t-online.de when being a user and wanted to redirect us to another service, we opted for a brief email discussion without any result.
\item[Wikipedia:] The response email forwarded the authors to their profile page, where a machine-readable copy might be downloadable. As this was the reason for the GDPR request, we decided against counting it as fulfilled.
\item[Zoom:] The textual answer received from Zoom includes information about deletion requests; unsubscribing from email marketing; access, rectification, and portability; and opt-out of sales (California residents). For accessing one's own data, Zoom forwards the reader to their profile.
\end{description}

\paragraph{Summary}

When applying the criteria of complete, user-friend\-ly, and machine-readable, we notice that only Facebook, Instagram, Snapchat, and Zalando fulfill them completely. While Facebook and Instagram either allow understandable or machine-readable data, new requests can be made right afterward. Although TikTok states that it provides an understandable format, TXT is maybe not the best way to present data to users. The other websites did not comply with at least one or more criteria, whereas 11/27 offered understandable data.

\subsubsection{Email-based vs. Website-based Request}
\label{sec:email}

When comparing email-based and website-based (browser website, app, or chat) requests, we noticed the following differences, as summarized in Table~\ref{tab:email} (MR: machine-readable, U: understandable).

\begin{description}
\item[Steps to request:] Generally, fewer steps (avg. 4.1, std. 1.5) are required to find the corresponding email address rather than to request data via the embedded function. On the other hand, the user has to copy a GDPR request template and adapt it for their purpose. Hence, it might reduce the likelihood of requesting data.
\item[Verification of request:] Email address and information about the account are suitable for the request.
\item[Time till answer:] For email cases, it took significantly longer (374.2 h avg.) to answer than otherwise (69.7 h avg.).
\item[Form of answer:] The data is sent via email and might be secured with a password or OTP.
\item[Conformity with GDPR:] Regarding the selected websites we ana\-lyzed, 25\% provided understandable and 12.5\% machine-readable data, none both. Hence, email-based requests perform significantly (p-value 0.002) worse than others.
\end{description}

In the methodology (see Section~\ref{sec:methodology}), we restricted the number of requests for websites that only offer email-based requests, as we assumed that they compile the data manually. Based on the observations, we conclude that our assumption was correct. To the authors, this seems surprising after five years of GDPR, although there might not have been enough requests to make an automated collection worthy. During the period of study but after our requests, one website -- BILD.de -- changed from email to a web form.

\begin{table}[!htpb]
\caption{Overview of the requests based on the type of request}
\label{tab:email}
{\def\arraystretch{1.1}
\begin{tabular}{lll}
\toprule
 \textbf{Request}                                                         & \textbf{Time}     & \textbf{Conformity}                                                                               \\ \midrule

 Email & \begin{tabular}[c]{@{}l@{}} 374.2 h avg.\\ 228.3 h std. \end{tabular}& \begin{tabular}[c]{@{}l@{}}MR: 12.5\%\\ U: 25\%\\ both: 0\%\end{tabular} \\ \midrule

 Website & \begin{tabular}[c]{@{}l@{}} 69.7 h avg.\\ 106.3 h std. \end{tabular}& \begin{tabular}[c]{@{}l@{}}MR: 57.9\%\\ U: 47.4\%\\ both: 31.6\%\end{tabular} \\ 

\bottomrule
\end{tabular}
}
\end{table}

\subsubsection{Comparison according to the Differentiations}
\label{sec:differentiations}

Further, we evaluate our data based on the differentiations \emph{a)}--\emph{d)}. In the following, we combine the discussion of differentiations \emph{a} (smaller and bigger organizations) and \emph{c} (location of the website).

We found no difference between browser website- and smartphone app-based usage, concerning the number of steps, type of answer, or duration from request until notification of the download. However, we noticed other differences.

\begin{itemize}
\item Changes to the app or web page can result in inconsistent user experiences concerning browser-to-app-based usage and different app versions.
\item Apps with a specific service, such as Amazon Music, might not have the request option included. Also, in other apps, it might be necessary to switch to the browser version for the request (see, for example, Reddit and Telegram).
\item Notification of the download is partly only in-app if the request was sent in-app. Hence, it might be missed. In contrast, a request via a browser-based website typically triggers an email notification. This behavior seems arbitrary, as some apps started to send advertisement emails after the request.
\end{itemize}

We found the following differences related to the location of the website, as summarized in Table~\ref{tab:location} (MR: machine-readable, U: understandable).
\begin{itemize}
\item Websites based in Germany (smaller audience) mainly prefer email requests (85.7\%) and answers (80\%), whereas otherwise website functions are applied (10.5\% email, 89.5\% browser-based website / smartphone app). In consequence, the level of identification is lower (email address and information) and the time from request to answer is higher (318.6 h avg., 246.5 h std.). In addition, conformity with GDPR (machine-readable: 28.6\%; understandable: 28.6\%; both: 14.3\%) is significantly lower than the conformity of worldwide websites (p-value 0.004).
\item Both Chinese websites offer machine-readable data, but there is no understandable version with a bigger variety of steps (4 and 10). In both cases, smartphone app or website functions lead to a comparable fast answer (24 h avg.).
\end{itemize}

\begin{table}[!htpb]
\caption{Overview of the requests based on the location}
\label{tab:location}
{\def\arraystretch{1.1}
\begin{tabular}{llll}
\toprule
\textbf{Location} & \textbf{Request}                                                         & \textbf{Time}     & \textbf{Conformity}                                                                               \\ \midrule
Germany           & \begin{tabular}[c]{@{}l@{}}Email: 85.7\%\\ Website: 14.3\%\end{tabular} & \begin{tabular}[c]{@{}l@{}}318.6 h avg.\\ 246.5 h std. \end{tabular}  & \begin{tabular}[c]{@{}l@{}}MR: 28.6\%\\ U: 28.6\%\\ both: 14.3\%\end{tabular} \\ \midrule
China             & \begin{tabular}[c]{@{}l@{}} Email: 0\%\\Website: 100\% \end{tabular}                                     & 12 h avg.                    & \begin{tabular}[c]{@{}l@{}}MR: 100\%\\ U: 0\%\\ both: 0\%\end{tabular}           \\ \midrule
Worldwide         & \begin{tabular}[c]{@{}l@{}}Email: 11.1\%\\ Website: 88.9\%\end{tabular} & \begin{tabular}[c]{@{}l@{}} 117.8 h avg.\\ 166.3 h std. \end{tabular} & \begin{tabular}[c]{@{}l@{}}MR: 44.4\%\\ U: 50\%\\ both: 27,8\%\end{tabular}    \\ \midrule
All               & \begin{tabular}[c]{@{}l@{}}Email: 10.5\%\\ Website: 89.5\%\end{tabular} & \begin{tabular}[c]{@{}l@{}} 156.7 h avg.\\ 204.6 h std. \end{tabular}& \begin{tabular}[c]{@{}l@{}}MR: 44.4\%\\ U: 40.7\%\\ both: 22.2\%\end{tabular} \\  
\bottomrule
\end{tabular}
}
\end{table}

The high percentage of email requests in Germany might be caused by the size of the organization or the number of users. In consequence, it has to be further evaluated in future studies.

Based on Table~\ref{tab:accounts}, we regard the three categories of social media, e-commerce, and others.
\begin{description}
\item[Social media:] The websites vary between four and ten steps, with 5.9 avg. and 2.1 std. None applies an email as the place for the request. Typically, authentication is good enough to request data with two exceptions, which require a password. On average, the requester receives their data quickly (17.1 h avg., 21.1 h std.). The download of the data is within the app or webpage with authentication. The data received is more likely better machine-readable (60\%) than understandable (50\%). In 40\% of cases, it is even both.
\item[E-commerce:] It takes on average 6 steps (std. 3.1). Email requests are applied in two cases (both German websites), with verification ranging from providing information to being authenticated and providing an access code or link. Depending on the expected level of automation, the request is either answered fast or slowly (166.4 h avg., 38578.6 variances, 196.4 h std.). To access the data, at least some authentication is required, with two cases requiring a higher level of security. The data is more often machine-readable (66.7\%) than understandable (50\%), with 33.3\% of websites providing both.
\item[Others:] The number of steps taken for the request is on avg. 5.3, std. 2.4. Both ways -- email (54.6\%) and website (45.5\%) -- can be found. To request the data, either an authentication to the website or information within the email of the request is needed. To access the data, either authentication or a password are required. This is possible on average after 259.9 hours (223.8 std, 50068.3 var.) and significantly (p=0.01) worse than the duration of social media responses. The data is less likely to be machine-readable (18.2\%) than understandable (27.3\%), with 0\% of the websites offering both.
\end{description}

The differences in of understandable data are not significant, in contrast to the duration from request to answer.

\section{Patterns}
\label{sec:darkpattern}

Dark patterns generally refer to design elements in websites, apps, or other interfaces that intend to mislead or manipulate users into making unintended choices or actions. According to Mathur et al.~\cite{DBLP:journals/corr/abs-2101-04843}, dark patterns have the attributes of asymmetry, covertness, deception, information hiding, restriction, and disparate treatment. Known dark patterns concerning GDPR include hiding opt-out options, using misleading language, or making it difficult to find the wanted information~\cite{10.1145/3313831.3376321,10.1145/3419249.3420132,10.1145/3511265.3550448,10.1145/3411764.3445779,10.1145/3359183}. Hence, these relate to cookie banners and consent. Based on the results, we find the following patterns concerning GDPR Article 15 requests. These might be without ill intent; nonetheless, they reduce the likelihood of requesting or receiving data.

\paragraph{Hiding GDPR request information}

The variations in steps to request their data may indicate that the functionality might be hidden. Here, we have to further differentiate between local steps (although many) or following page by page, reading long texts to find the next link. Whereas many steps may discourage the user, this is clearly the case if the user path is unclear. The latter happened, for example, at AliExpress, where we had to go through at least three long web pages to proceed. Interestingly, searching for GDPR either on the website or with search engines did not help in all cases, for example, if no result is found on the website or the cached page is unavailable without any redirect.

\paragraph{Changing the place for GDPR requests}

Changing the place for GDPR requests without updating guidelines and adding redirects is another pattern. In some cases, such as with Twitter, we tried to apply the available guidelines without any result. Here, we also noticed differences between browser-based and app-based websites, as well as different app versions. Hence, the user does not have a consistent experience, which might make future requests unlikely if the corresponding web page is not found at the known location. 

\paragraph{Inconsistent GDPR request options}

Bigger website providers accumulate several services. Whereas most had one place to request their data, this was not always the case. For example, Twitch, an Amazon service, uses a different request type than Amazon itself. The worst in this respect is according to our experience with Microsoft. While searching for a way to request our Skype data, we found the Microsoft privacy dashboard by utilizing a search engine. Here, we were forwarded to the website's corresponding page, which had additional information and forwarded us further. Curiously, we tested other websites with different results, ranging from email to forms and no possibility at all.

\paragraph{Making it impossible to access GDPR request}

Corresponding to the first and second patterns, we noticed broken and missing links, while searching for the according information. This was more likely in the email cases. If a Uniform Resource Locator (URL) changes, then the corresponding links should be updated and redirects should be established, guiding the user to the new location. Missing links, such as in the case of eBay or Chefkoch, put another burden on the user to make a request. For example, the contact information was said to be in contact without a link to the contact section. We did not find the corresponding information and proceeded through further loops.

\paragraph{Requiring too much or no proof of identification}

In order to send the request and acquire the data, the user might need to prove their identity. For email requests, this was basically the username and email address of the account, which could be gathered otherwise. Two exceptions exist: the password for encrypting the data was sent to the verified phone number, or the password was sent in a second email. Otherwise, no verification was needed. For website-based requests (i.\,e., browser, smartphone app, chat), it was mostly enough to be authenticated. In a few cases, passwords, tokens, or re-authentication were required. These extra steps might be hidden, for example, at AliExpress. In consequence, adversaries might be able to receive the data if they know the required information or can authenticate. In contrast, too many steps or too high proof (such as a copy of an ID card) might reduce the number of users actually submitting such a request or accessing their data.

\paragraph{Missing notifications}

Whereas web browser requests generated an email notification and email cases were answered by email, app requests did not always result in an email, informing the user that the data export was ready to download. If no notification within the smartphone's notification area is displayed and the user is not using the app daily, then the time frame for downloading the data can be missed. In addition, in some apps, such as Snapchat, the notification is not visible when opening the app, but hidden in the menu. In consequence, the user might also miss the download.

\paragraph{Making it impossible to access data}

Not only proofs, but also technical functionalities may make it impossible to access the data. For the app case, an older smartphone was used. Consequently, the speed of some actions was reduced, which could also be the case for elderly or (temporarily or situationally) disabled persons. For Twitch, it was impossible to solve the captcha within the set time frame, making it impossible to download the data. Here, we had to switch to the browser version. In one case, the page for the download was hidden and could only be accessed by clicking on the link in the notification or email. Hence, if the notification is deleted or not displayed, the user cannot access their data.

\paragraph{Not answering GDPR request (correctly)}

As explained above, not all websites answer the GDPR Article 15 request or send inadequate data/answers (according to Articles 5, 12, and 15, mainly). In one case, we had an email discussion with the responsible person about whether they had to provide the data for their website, as they forwarded us to another website that we were clearly not using. Furthermore, we noticed that two websites do not allow more than one request within a month or longer. In many cases, the data is not understandable.

\paragraph{Sending data that is not linked to the user}

To display understandable data in the form of a privacy dashboard to the user, additional data, such as fonts, icons, and JavaScript, might be embedded. Although these enable a similar user experience to a normal website, the files might be misunderstood by users with no further knowledge about web development. If additional data is applied, then the HTML entry page is located in the main folder, while the other data is included in subfolders. Hence, the user is directed to open the HTML page, which is only possible if the ZIP archive is extracted. With no guidelines, some users might end up not actually seeing their data.

\paragraph{Hiding data from the user}

Last but not least, in the case of Twitter, the user-friendly privacy dashboard might not include all data, as explained in the \texttt{README.txt} file located between the JavaScript files. In contrast, the JavaScript files contain all the data, according to Twitter. As a result, the user might be fine with the data displayed, not knowing that Twitter has more (serious) data about them -- even though it was sent to them.

\section{Discussion}
\label{sec:discussion}

We noticed no differences in conformity depending on the smartphone app- or browser-based usage, but on the type of request, i.\,e., email or otherwise. This may correlate with the location and size of the organization or the number of users they have. Additionally, we analyzed the generic groups of social media, e-commerce, and others with comparably high conformity for social media and e-commerce. This provides a first indication of different behavior depending on the type of website. For verification, it requires further research with the inclusion of additional websites and more fine-grained categories. Most websites handle GDPR Article 15 requests, although not all comply with completeness  (Article 5 (1)) and understandability (Article 15). Often, the data was machine-readable (Article 20 (1)) instead. The biggest influencing factor we observed is automation, see email- vs. website-based requests and answers. For our analysis, we mainly sent the request after one month of usage, as some user data is stored at that time. Nevertheless, for automated requests, we also measured the time between the request and the answer for the initial request (see Article 12 (3)). We noticed that Google (8h vs. 11h) and Dropbox (26h vs. 87h) required more time at the second request, which was not the case for Amazon (166h vs. 107h) -- although Amazon states that depending on the amount of information in the account, it might take up to a month. Hence, the age and included data of an account might result in a longer processing time. As shown in Section~\ref{sec:process}, the burden of identification (Article 12 (6)) is almost nonexistent. Hence, it is comparably simple to request one's data if the steps to find the functionality are not too high a barrier. At the same time, this also increases the possibility for adversaries to request and access data. On the other hand, if an adversary already has access to an account, they could alter it. The selection of PayPal was omitted due to the required proof of identification (copy of ID). In Section~\ref{sec:darkpattern}, we describe ten observed patterns.  While not fully complying with the GDPR might be obvious, see also related work in Section~\ref{sec:sota}, other patterns, such as making it impossible to access the data or confusing the user experience, are not. 
During our qualitative study, selected apps changed the steps required to request a data extract. These changes might stop users from requesting their data if they cannot find the functionality in the known location. This might be of less importance if the workflow is still intuitive and requires fewer steps. Based on this observation, a long-term study would be interesting, recognizing all the changes over the years. During the registration of the accounts for the synthetic user, we decided on a non-concerned user according to the privacy attitude of Zhang et al.~\cite{10.1145/1081870.1081913}. In consequence, the user applied Single-Sign On (SSO) with Google and accepted all privacy settings and notifications as presented by default. Due to SSO, one might expect that the account needs to be secured in a better way. Nonetheless, simple passwords were possible for Google and other websites.

\section{Comparison with Previous Studies}
\label{sec:sota}

Several authors address GDPR in general. Ausloos and Veale~\cite{Ausloos_Veale_2021} outline an approach for GDPR data request research and discuss ethical and methodological considerations. Arfelt et al.~\cite{10.1007/978-3-030-29959-0_33} focus on understanding data holders' conformity with legislation, whereas Zaeem and Barber~\cite{10.1145/3389685} analyze the effect of GDPR on privacy policies. Wong and Henderson~\cite{10.1145/3267305.3274152} discuss data portability. Our study underlines the problems described by these authors. In contrast, Liu et al.~\cite{10.1145/3442381.3450022} analyze the compliance with GDPR Article 13 and Mangini et al.~\cite{10.1145/3407023.3407080} the right to be forgotten. Several authors examine cookie banners and their usage~\cite{10.1145/3292522.3326039,10.1145/3466722,10.1145/3308558.3313524,10.1145/3321705.3329806}, GDPR consent~\cite{10.1145/3442381.3450056,10.1145/3313831.3376321,10.1145/3548606.3560564}, and dark patterns related to it~\cite{10.1145/3313831.3376321,10.1145/3419249.3420132,10.1145/3511265.3550448,10.1145/3411764.3445779,10.1145/3359183}. 

In addition, some approaches consider GDPR access to data either from a privacy threat perspective or through user studies. Spagnuelo et al.~\cite{icissp19} review transparency-enhancing technologies to accomplish transparency in accordance with the GDPR. Furthermore, \cite{9283991,10.5555/3361476.3361504,Clarke2019} consider the threat to privacy through insufficient ID verification. The observations are in accordance with our study. Alizadeh et al.~\cite{10.1145/3340764.3344913} conducted a study with 13 users of a German loyalty program, finding better responses and education were required. Similarly, Pins et al.~\cite{10.1080/0144929X.2022.2074894} recruited 59 participants to exercise their rights to access their data, with comparable results. Bowyer et al.~\cite{10.1145/3491102.3501947} assess the practical experiences of GDPR data requests in a 10-participant study, where each participant filed four to five data access requests. Also here, the request for understandable data (Article 15) was not met. The authors propose a more effective design based on the results. Although the user perspective is outlined, none of these approaches analyze GDPR requests and responses from a scientific perspective. Meaning that no clear test cases were applied to several websites to analyze their compliance with GDPR Article 15 within a study. In addition, our results underline the existence of patterns concerning the implementation of Article 15. However, this requires further investigation in future work.

\section{Conclusion and Outlook}
\label{sec:conclusion}

The EU GDPR was implemented five years ago to strengthen and harmonize the data protection of individuals. One article, Article 15, gives individuals the right to access their personal data. Consequently, organizations had at least five years to optimize their processes and functions to comply with Article 15, among others. This study aimed to explore the process of requesting and receiving the responses of organizations to GDPR Article 15 requests. In our qualitative analysis involving a natural and an artificial user applying both browser-based and app-based websites, we obtained data from those selected websites to analyze the reaction to GDPR Article 15 responses and possible patterns. We thereby followed the methodology outlined in Section~\ref{sec:methodology} based on the GDPR (see Section~\ref{sec:gdpr}) to gain our results, as described in Section~\ref{sec:evaluation}, divided into workflow and data analysis. Based on the results, we summarize the observed patterns in Section~\ref{sec:darkpattern}. The results were then discussed in Section~\ref{sec:discussion} and compared with the literature in Section~\ref{sec:sota}. In a long-term study, we want to measure changes and improvements related to reactions to GDPR requests and the incorporation of the data hinted at in the privacy statements. We aim to analyze various websites within different countries and privacy compliance outside the GDPR. Last but not least, we plan to conduct a user study to better understand the usage of the data requests.

\bibliographystyle{ACM-Reference-Format}
\bibliography{gdpr}

\end{document}